\documentstyle[prb,aps,epsf]{revtex} \def\narrowtext{} \tighten \twocolumn
\input epsf.sty

\begin{document}
\draft

\title{Influence of vortices on the magnetic resonance in cuprate
superconductors}

\author{M. Eschrig$^{a}$, M. R. Norman$^{a}$,and B. Jank{\'o}$^{a,b}$}
\address{ $^a$Materials Science Division, Argonne National Laboratory, Argonne,
Illinois 60439}
\address{ $^b$Department of Physics, University of Notre Dame,
Notre Dame, Indiana 46556}
\address{%
\begin{minipage}[t]{6.0in}
\begin{abstract}
We investigate several theoretical possibilities for the suppression in a
c-axis magnetic field of the magnetic resonance recently observed in inelastic neutron
scattering experiments on YBa$_2$Cu$_3$O$_{6.6}$.
We find that neither the Doppler shift of the quasiparticle states caused by 
supercurrents outside the vortex core, nor an assumed
spatially uniform suppression of the coherence factors or spectral gap due
to the applied field, can account for the observed effect.
In contrast, suppressing the gap or the coherence factors in the vortex core
to zero is consistent with the data.
We demonstrate that an even simpler description of the data can be achieved
by assuming that the resonance is not supported within
an effective radius $\xi_{eff}$ around each vortex, 
where $\xi_{eff}$ is the sum of the superconducting
and spin-spin correlation lengths.  We use this simple idea to predict the
doping dependence of the field suppression.
\typeout{polish abstract}
\end{abstract}
\end{minipage}}

\maketitle
\narrowtext

One of the more intriguing developments in the field of high temperature
cuprate superconductivity has been the observation by inelastic neutron
scattering (INS) experiments of a sharp magnetic resonance in the 
superconducting state\cite{RES}.  Recently, it was found that a 
c-axis magnetic field suppressed the intensity of this resonance\cite{Dai00},
as predicted from an analysis of specific heat data\cite{Janko99}.
Since the same effect was not observed for in-plane fields\cite{Bourges},
this indicates that the resonance is sensitive to the presence of Abrikosov
vortices, and thus intimately connected to the nature of the superconducting
ground state.  This has obvious implications for microscopic theories of
the resonance. 

In this paper, we consider a model where the resonance is treated as a
particle-hole bound state in a $d$-wave superconductor, with calculations
performed within linear response theory (RPA).  Several effects of the vortices
are considered.  First, we calculate the influence of the supercurrents
circulating around the vortices on the resonance.  We find that this only
leads to a broadening of the resonance in energy; the integrated weight remains
the same, in conflict with experiment.  Second, we study the effect of a
spatially uniform suppression of the 
$\langle \Delta_{{\bf k}}\Delta_{{\bf k}+{\bf Q}} \rangle $ correlator
which enters the coherence factors of the spin susceptibility
(where ${\bf Q}$ is the antiferromagnetic wave vector at which the resonance is
peaked).  Such a suppression is speculative, but could be a result of
dephasing of the pairing in a c-axis field due to the vortices, as observed
in Josephson plasma resonance experiments\cite{JOS}.  We find that
although this does lead to a suppression of the integrated weight as
observed experimentally, the effect causes the resonance
to shift to higher energy, in conflict with experiment.  Third, an assumed
(field induced) spatially uniform suppression
of the gap magnitude causes the resonance to shift to lower energy, also
in conflict with experiment.

This leads us to consider the effect of the vortex cores themselves.
We observe that if the resonance is not supported in the vortex cores,
then the resulting field dependence is in reasonable agreement with
experiment.  
We consider three possibilities for the suppression of the resonance in the
vortex core regions:  (a) the suppression of the gap magnitude in
the core, (b) the suppression of the $\langle \Delta \Delta \rangle $ correlator
in the core, and (c) the absence of quasiparticles in 
the core.  Any of these three possibilities give a good account of the data.
We use case (c) to estimate the doping
dependence of the field suppression effect.

To calculate the influence of the supercurrents around the vortices on the
resonance in the spin-spin correlation function, we approximate the superflow
by a circular flow around the vortex center. The corresponding local
supermomentum, ${\bf p}_s $, is proportional to the gradient of the phase,
${\bf p}_s=\hbar {\bf e}_\phi/2r $. This is a good approximation for the experiments
considered here,
where the intervortex spacing is smaller than the penetration
depth and large compared to the coherence length.
 
In the intervortex regions, the variation of
the order parameter and of the superflow occurs on a scale large compared
to the spin-spin correlation length, which amounts to only a few lattice
constants as determined from the momentum width of the resonance.
Consequently, we
determine the RPA susceptibility in the intervortex region at each point of the
unit cell of the vortex lattice in the presence of the {\it local} superflow,
\begin{equation}
\chi (\omega, {\bf Q},{\bf p}_s)= \frac{\chi_0(\omega ,{\bf Q},{\bf p}_s)}{1-J_{{\bf Q}}
\chi_0(\omega ,{\bf Q}, {\bf p}_s)} \; .
\label{chi}
\end{equation}
The bare susceptibility $\chi_0(\omega ,{\bf Q},{\bf p}_s)$, is determined as
\begin{eqnarray}
\chi_0(\omega ,{\bf Q},{\bf p}_s) &=& -\sum_{{\bf k}} \sum_{\mu,\nu =\{\pm \}}
\frac{A_{{\bf k}}^\mu A_{{\bf k}+{\bf Q}}^\nu + \alpha C_{{\bf k}}^\mu C_{{\bf k}+{\bf Q}}^\nu }{\omega + 
E_{{\bf k}}^\mu-E_{{\bf k}+{\bf Q}}^\nu+i\Gamma } \nonumber \\
&&\times \left(f(E_{{\bf k}}^\mu ) - f(E_{{\bf k}+{\bf Q}}^\nu )\right)
\label{chi0}
\end{eqnarray}
where the excitation spectrum in the presence of a superflow with momentum
${\bf p}_s $ is given by\cite{deGennes}
\begin{eqnarray}
E_{{\bf k}}^\pm &=& \pm \sqrt{ \bar \xi_{{\bf k}}^2
+|\Delta_{{\bf k}} |^2} +\delta \xi_{{\bf k}}
\end{eqnarray}
with $\delta \xi_{{\bf k}} = (\xi_{{\bf k}+{\bf p}_s }-\xi_{{\bf k}-{\bf p}_s })/2 $,
$\bar \xi_{{\bf k} }= (\xi_{{\bf k}+{\bf p}_s}+\xi_{{\bf k}-{\bf p}_s})/2 $, and
\begin{eqnarray}
A_{{\bf k}}^\pm = \frac{1}{2}\pm \frac{\bar \xi_{{\bf k}} }{E_{{\bf k}}^+-E_{{\bf k}}^-}  
\qquad C_{{\bf k}}^\pm = \pm\frac{ \Delta_{{\bf k}} }{E_{{\bf k}}^+-E_{{\bf k}}^-}
\end{eqnarray}
as coherence factors. The factor $\alpha $ (1 in the current case)
will be discussed later.
The spatial average of the Doppler shifted susceptibility over the
intervortex region of the
vortex lattice unit cell is then calculated,
$\chi (\omega ,{\bf Q})=<\chi ( \omega ,{\bf Q} ,{\bf p}_s({\bf R}) ) >_{{\bf R}}$.  
We evaluated Eq.~(\ref{chi0}) for a $512\times 512 $ grid of
${\bf k}$-points and perfomed the spatial average over 320 ${\bf R}$-points.
Note we use a normalization equal to the intervortex area, not the total area.
The contribution from the vortex cores will be discussed later in the paper.

Calculations were performed using a model quasiparticle dispersion in the
superconducting state motivated by photoemission measurements\cite{NE}.
Similar dispersions were found to give a good description of the zero field INS
data, including the incommensurate structure observed at energies below
resonance \cite{NORM01}.  A $d$-wave superconducting gap 
proportional to $\cos(k_xa)-\cos(k_ya)$ 
was assumed, with a maximum value of $\Delta =$ 29 meV as
determined from recent STM measurements\cite{WEI}.  A broadening factor,
$\Gamma$, of 2 meV was employed, and a temperature of 13 K.

\begin{figure}
\centerline{
\epsfxsize=0.50\textwidth{\epsfbox{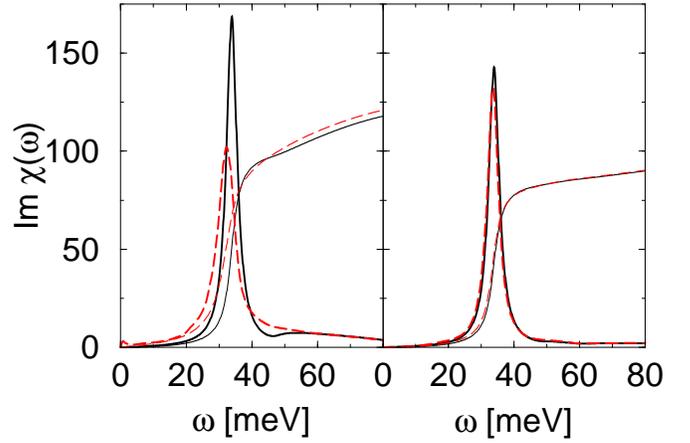}}
}
\caption{ 
\label{fig1}
Influence of Doppler shifts due to supercurrents on Im $\chi (\omega ,{\bf Q}) $. 
The full lines correspond to zero magnetic field, the dashed ones to a field
of 7 T.
The thin lines are the energy integrated susceptibilities (scaled by
a constant so that they could be plotted on the same graph).
Although the magnitude of the resonance at the peak energy is suppressed, its
integrated weight is not.
The parameters employed were 
$\Delta$=29 meV, $\Gamma$=2 meV, $T$=13 K.
In the left panel we use $J$ = 357 meV, and the dispersion
taken from Ref.~\protect\onlinecite{NE}. In the right panel 
we use $J$ = 142 meV and the dispersion $tb2$ taken from
Ref.~\protect\onlinecite{NORM01}; for this case, the Doppler shift has
virtually no effect.
}
\end{figure}

In Fig.~\ref{fig1} we show our results for the effect of the circulating
supercurrents on the resonance.  The exchange coupling, $J_{{\bf Q}}$, is
fixed to give a resonance at 34 meV for zero magnetic field.  For the
spatial average we assumed a lower cut-off at the value $\xi=2a$
(vortex core radius) and an upper cut-off at the value $R=25 a$
(radius for enclosing one flux quantum at 7 T), where $a$ is the Cu-Cu
distance.  The results are insensitive to the lower cutoff.  As Fig.~\ref{fig1}
shows, the supercurrent has three effects: (a) 
it shifts the position
of the resonance to slightly lower energy, (b) it broadens the
resonance, and (c) it reduces the magnitude of the resonance at the peak
energy.  Also shown are the energy integrated susceptibilities, which
demonstrate that the integrated weight between 0 and $\approx 2\Delta$
is conserved.  These findings are in apparent contradiction with the
experimental facts, which are that the resonance does not shift, nor
broaden, and that the integrated weight is reduced by about 15\% at 7
T [Ref. ~\onlinecite{Dai00}].  We have also tested a number of other
dispersions \cite{NORM01}, and a variety of assumed values for
$\Delta$ and $J$.  Although the amount of broadening is somewhat
sensitive to these details, we find that the integrated weight is
always approximately conserved.  
An example is given in the right panel of Fig.~\ref{fig1}, where we find
virtually no effect of the Doppler shift on the susceptibility.  

\begin{figure}
\centerline{
\epsfxsize=0.45\textwidth{\epsfbox{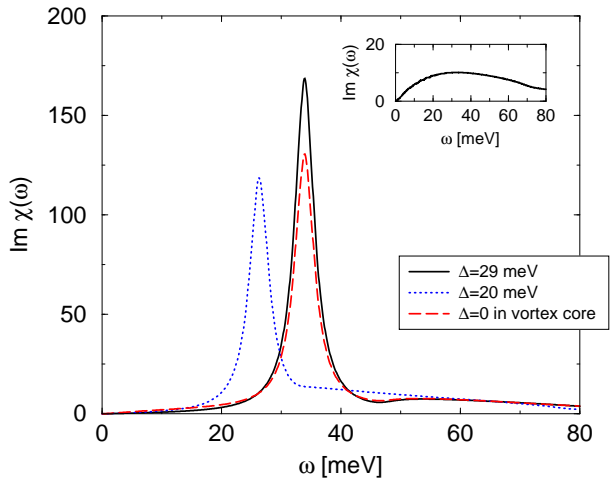}}
}
\caption{
\label{fig2a}
Comparison of zero field susceptibility with the same susceptibility, but
with reduced gap magnitude, $\Delta$.
Dotted line: assuming a spatially uniform reduction from 29 meV to 20 meV;
dashed line: assuming a reduction to zero in 24\% of the
vortex unit cell area representing the cores (the uniform zero gap
response is shown in the inset).
In both cases the weight of the resonance is reduced, but
for the spatially uniform case, the resonance is shifted considerably
downwards in energy.
The parameters used are the same as in Fig.~\ref{fig1}.
}
\end{figure}

We also checked if an assumed field induced (spatially uniform)
reduction of the gap magnitude
accounts for the observed effect. Our result is shown as the dotted line
in Fig.~\ref{fig2a} compared to the zero field result (full line). 
The integrated weight is suppressed in this case
(the left panel 
of Fig.~\ref{fig3} shows the reduction of the integrated weight
versus $\Delta^2(H)/\Delta^2(0)$).
To obtain the observed 15\% 
\noindent
reduction in weight at 7 T would
require reducing the gap from 29 to 20 meV.
This reduction is substantially larger than would be indicated by the upper
critical field (45 T), and
the reduction appears to have the wrong functional dependence on $H$.
Moreover, this gap reduction shifts the resonance to considerably lower energy,
in contradiction with experiment.

As a third mechanism, we studied a (spatially uniform) suppression of the
$\langle \Delta_{{\bf k}}\Delta_{{\bf k}+{\bf Q}} \rangle $ correlator in the 
$C_{{\bf k}}C_{{\bf k}+{\bf Q}}$ coherence
factors (by reducing $\alpha$ to less than 1 in Eq.~\ref{chi0}).
The motivation for this is that phase fluctuations induced by the
vortices are known to lead to a dephasing of the layers, and the
resonance will be sensitive to this since it involves c-axis coupling
(it is peaked at $k_z=\pi/d$, where $d$ is the separation of nearest
neighbor CuO layers).
The observed decoupling inferred from the field
dependence of the Josephson plasmon \cite{JOS}, though, is probably
due to the weaker bilayer-bilayer coupling, which is also consistent
with small mesa experiments \cite{KLML}.  Therefore, at the current
time, it is not known whether the two layers within a bilayer are
dephased or not (though this could be determined from the field
dependence of c-axis infrared conductivity measurements, where a
feature is seen attributed to an optical Josephson plasmon \cite{VDM}). 
For now, though, we will assume that this is a
possibility, and test its consequences \cite{IM}.

\begin{figure}
\centerline{
\epsfxsize=0.45\textwidth{\epsfbox{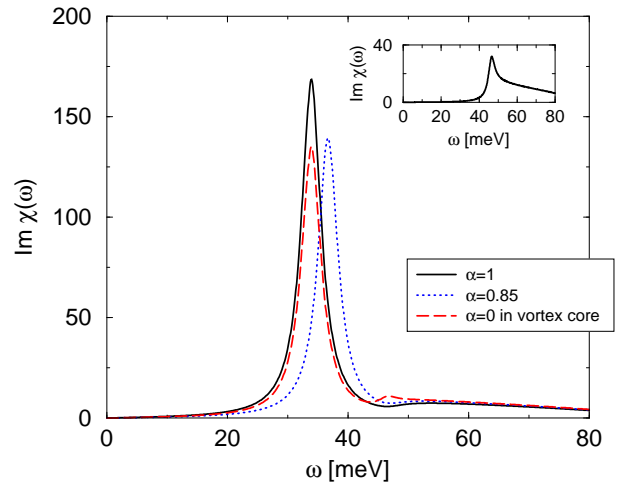}}
}
\caption{
\label{fig2b}
Comparison of zero field susceptibility with the same susceptibility, but
with the $\langle \Delta \Delta \rangle $ correlator in the numerator
of Eq.~2 reduced ($\alpha < 1$). Dotted line:
assuming a spatially uniform reduction by 15\%;
dashed line: assuming a reduction to zero in 20\% of the
vortex unit cell area representing the cores (the uniform
zero $\alpha $ response is shown in the inset).
In both cases the weight of the resonance is reduced, but
for the spatially uniform case, the resonance is shifted considerably
upwards in energy.
The parameters used are the same as in Fig.~\ref{fig1}.
}
\end{figure}

\begin{figure}
\centerline{
\epsfxsize=0.24\textwidth{\epsfbox{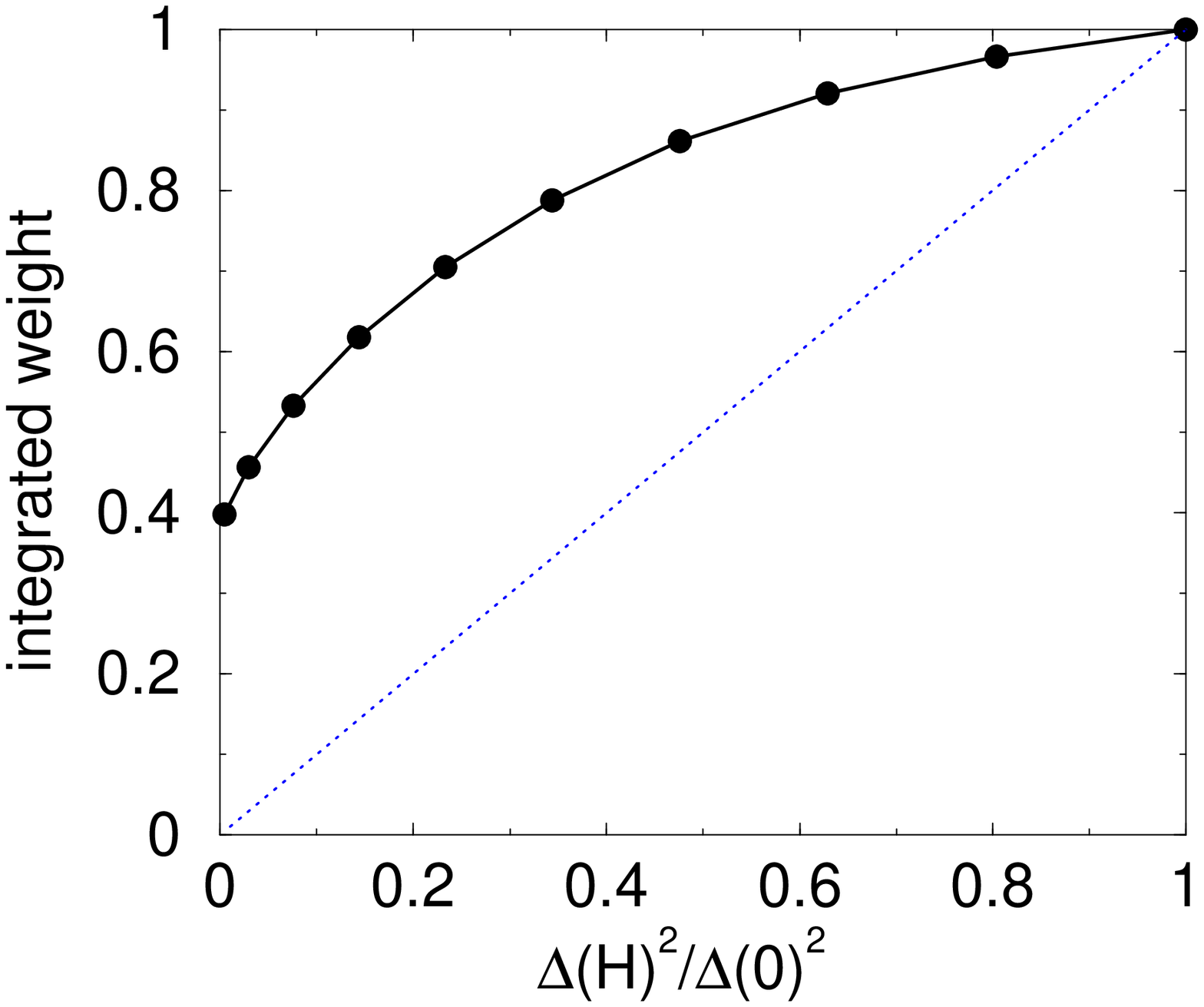}}
\epsfxsize=0.24\textwidth{\epsfbox{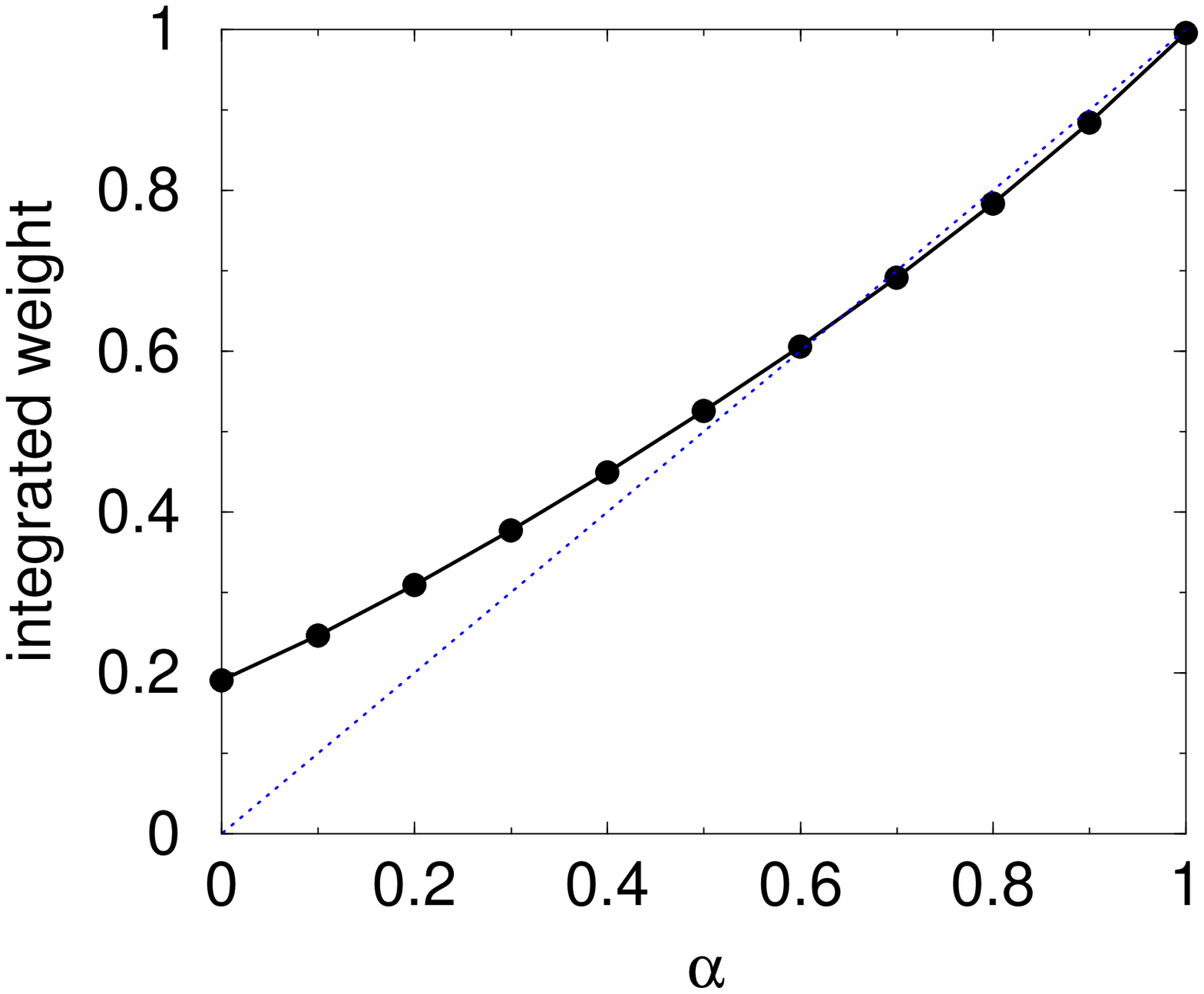}}
}
\caption{
\label{fig3}
Energy integrated weight (from $\omega=0$ to 
$\omega=50$ meV, normalized to the zero field value) from Figs.~\ref{fig2a}
and \ref{fig2b}.
Left: versus 
$\Delta^2(H)/\Delta^2(0)$, where 
$\Delta$ is the maximal $d$-wave gap magnitude. 
Right:
versus $\alpha$, where $\alpha$ is the
prefactor in front of the $\langle \Delta \Delta \rangle $ correlator in the numerator
of Eq.~2.  The dotted lines are 
the expected behavior if the normalized weight
was equal to $(1-H/H_{c2})$.
}
\end{figure}

In Fig.~\ref{fig2b}, we compare the zero field result to the same result, 
but with the correlator reduced by 15\% ($\alpha=0.85$).  
This leads to a large reduction of the
integrated weight, as seen experimentally (in Fig.~\ref{fig3}, 
we plot the integrated weight versus $\alpha$).
We note that the 
\noindent
experimental suppression goes like $1-H/H^*$, where $H^*$
is a number not much lower than $H_{c2}$, the upper critical field \cite{Dai00}.
Based on quantum Ginzburg-Landau theory, a reduction of the $\langle \Delta \Delta \rangle $
correlator proportional to $1-H/H_{c2}$ is expected.
Therefore, it is reasonable to suppose that the relative experimental
suppression goes like $\alpha$.  This suppression is in good agreement with
the calculation, as can be seen in Fig.~\ref{fig3}.  
We note, however, that the position of 
the resonance shifts to higher energies, in disagreement with the data.
It would be coincidental if this energy shift was exactly canceled by an
assumed shift of the superconducting gap to lower energies by the field.

Let us now consider the effect of the vortex cores.  The
fact that the experimental suppression goes like $1-H/H^*$ is highly suggestive
of a vortex core effect, as originally noted by Dai {\it et al.} \cite{Dai00}.
Therefore, we assume that the resonance is not supported in the region of the
vortex core. This assumption is based on five facts: 
a) the considerable momentum
width of the resonance shows that the corresponding spin excitations 
have a decay length of only two lattice constants, which is smaller than
the coherence length; thus the resonance will be sensitive to variations of the
order parameter on the coherence length scale; 
b) the resonance at zero field only exists in the
superconducting state, and disappears in the normal state;
c) coherence peaks in the single particle density of states at the gap edge
were not found in the core region in STM measurements \cite{STM};
this would modify the $2\Delta $-edge in $\chi''_0$ (Eq. \ref{chi0})
and suppress the resonance;
d) in underdoped materials, missing subgap states 
point towards a loss of quasiparticle 
weight due to a pseudogap in the vortex core; \cite{STM} 
e) the dip feature in the tunneling density 
\noindent
of states, thought to be due to the coupling of 
quasiparticles to the resonance \cite{NE}, is not observed in the
vortex core region \cite{STM}.

In Figs.~\ref{fig2a} and \ref{fig2b}, we show in the insets the susceptibility
for zero $\Delta $, and for zero $\langle \Delta \Delta \rangle $ correlator. In both
cases, the resonance is strongly suppressed.  In the main panels, we show as
dashed curves the results for the case when we use the curves in the insets 
for the vortex core regions, and the full curves (zero field results)
for the intervortex regions.
The latter is justified since we found above that the Doppler shift has a
negligible effect on the integrated intensity.
In both cases, the resulting curves, calculated for a 15\% reduction in
total integrated weight,
reproduce very well the experimental finding of no shift or broadening of
the resonance.
One problem is that the observed weight
suppression would require a vortex core region filling 24\% of the
total area for the first case, and 20\% for the second case.
Both values are somewhat larger than what one would expect at 7 T,
as discussed below, especially for the first case.

There is a way of testing whether either of these two scenarios is correct.
In the case where the gap is reduced in the cores, then extra weight
would show up at lower energies, whereas for the case where the correlator
is reduced in the cores, extra weight would show up at higher energies.
Although our calculations
indicate that the corresponding changes between 0 and 7 T are 
perhaps within experimental error bars, we encourage experimentalists
to look for extra weight, both in the region around 
20 meV and around 50 meV.

We note that our formalism assumes
the presence of quasiparticle states, and thus we do not expect a resonance
to exist if quasiparticles do not exist.  
A simpler idea to those discussed in the previous
two paragraphs is to simply assume that the
weight is zero in the considered energy range,
consistent with a spin gap of $\sim $ 50 meV inside the cores.
From this, we can estimate the effective 
radius of the core, $\xi_{eff}$, from the ratio of the INS
weight to that at zero field.
Since the ratio of the vortex core area to the total area is equal to
$B\pi \xi_{eff}^2/\Phi_0$, to obtain an effect of 15\% at 7 T in 
underdoped YBa$_2$Cu$_3$O$_{7-\delta }$, we need a $\xi_{eff}$ of 37-38 \AA.
This is somewhat larger than the estimated coherence length of 27 \mbox{\AA }
($H_{c2}$ of 45 T), 
but it should be noted that $\xi_{eff}$ should approximately be the sum of
the superconducting {\it and} magnetic correlation lengths.  The latter is
approximately 10 \mbox{\AA } 
(as measured by the momentum width of the resonance),
so the two estimates are in agreement.

For optimally doped compounds, both coherence lengths are shorter.
Assuming $\xi_{eff}\approx $20-24 \AA,
our prediction for this case is that the suppression of the total weight 
at 7 T will only be 4-6\%. Going further to the overdoped regime, the
superconducting coherence length increases, leading to an increase of
the sensitivity of the resonance with magnetic field again.  Further 
underdoping, though, should lead to an even more dramatic reduction, as 
both coherence lengths are expected to increase as the doping is reduced.
In fact, we would argue that the field dependence of the resonance at various
dopings would be a good measure of the doping dependence of the 
superconducting coherence length, and it would be of great interest to 
correlate STM and INS measurements on the same samples.

We wish to conclude this paper with the following speculation, motivated
by the above results.  As documented by
angle-resolved photoemission measurements \cite{arpes},
quasiparticle-like peaks in the spectral functions are present
only below $T_c$, the onset temperature of
phase coherence.  The superconducting phase is
singular at the vortex core, and therefore the phase correlations are
strongly suppressed between points close to the core region (this was a
motivation for the $\alpha < 1$ calculations).  
We suggest that this may lead to a destruction
of quasiparticle excitations in the vortex core region
similar to what happens in the pseudogap state.
The absence
of quasiparticle peaks as well as the neutron resonance in the core
region is consistent with the notion that {\em both} these spectral
features require substantial local phase correlations\cite{Janko99}.
While this conjecture is at this stage admittedly speculative, we
believe it deserves further experimental and theoretical investigation.

This work was supported by the U.S. Dept. of Energy, 
Office of Science, under Contract No.~W-31-109-ENG-38.

\end{document}